\documentclass{bms}
\usepackage{lipsum}
\usepackage{amssymb}
\usepackage{amsfonts}
\usepackage{mathtools}
\usepackage{graphicx,amsmath,bm}
\usepackage{cite}
\usepackage[utf8]{inputenc}
\begin{document}

%%paper title
%%For line breaks \\ can be used within title
\title{Application of k $\cdot$ p method on band structure of GaAs obtained through joint density functional theory}

%%author names are separated by comma (,)
%%use \and before the last author name
%%\textsuperscript{number} is used for affiliation
%%use a * along with the number separated by comma
%% for the  author for correspondence

%\author{Waqas Mahmood\textsuperscript{1, *} \and Bing Dong \textsuperscript{1}}
\author{Waqas Mahmood\textsuperscript{*} \and Bing Dong\textsuperscript{}}

\affilOne{\textsuperscript{}School of Physics and Astronomy, Shanghai Jiao Tong University, Shanghai, 200240 China\\}
%\affilTwo{\textsuperscript{2}School of Physics and Astronomy, Shanghai Jiao Tong University, Shanghai, 200240 China}

%%escape two column mode for title, affiliation and abstract
%%by giving \twocolumn command as shown

\twocolumn[{
\maketitle
%%include \msinfo for
%%manuscript information such as
%%received, revised and accepted dates
%%
%\msinfo{1 July 2015}{20 April 2016}
%%abstract
\begin{abstract}
The structural and electronic properties of zinc-blende (ZB) GaAs were calculated within the framework of
plane-wave density-functional theory (DFT) code JDFTx by using Becke 86 in 2D and PBE exchange correlation functionals from libXC. The standard optimized norm-conserving Vanderbilt pseudopotentials were used to calculate optimized lattice constant, band gap and spin-orbit split-off parameter. The calculated values of optimized lattice constant and direct band gap are in satisfactory agreement with other published theoretical and experimental findings. By including spin-orbit (SO) coupling, conduction band and valence bands were studied under parabolicity to calculate effective masses. The calculated values of effective masses and spin-orbit split-off parameter are in satisfactory agreement with most recent findings. This work will be useful for more computational studies related to semiconductor spintronic devices.
\end{abstract}
%%insert keywords separated by semicolon using \keywords{words}
}]
%%close the twocolumn escape here
%%include \doinum{number}for the DOI number in the header
%%include \volnum{number} for the volume number in the header
%%include \year{yyyy} for  year of publication in the header
%%include \pgrange{num--num} page range of article in the header
%%include \artcitid{num} for the article citation id
%%include \lp to print last page of the article
%%include \setcounter{page}{pagenum} for the exact starting page of the article
%%include \corres to print the corresponding author Email id
\corres{waqasmahmoodqau@sjtu.edu.cn}
%\runauthor{Short Author}
%\runtitle{Short Title}
%\doinum{XX.XXXX/XXXXXX-XXX-XXX-X}
%\volnum{39}
%\issuenum{7}
%\monthname{December}
%\year{2016}
%\pgrange{1625--1627}
%\setcounter{page}{1625}
%\lp{1627}
%\vspace{-36pt}
\section{Introduction}
The electron-spin interactions give rise to interesting semi-conductor spintronic effects and allow us to control charge and spin \cite{FabianDasSarma,FabianMatos}. In III-V semi-conductors, spin-orbit coupling manifests itself in band structure by splitting energy bands and preserving spin degeneracy as space inversion symmetry is not present in primitive cell of these semi-conductors \cite{CamposGmitra}. Lack of space inversion symmetry generates momentum dependent SO field in analogy to Zeeman field that splits energy bands. In ZB semi-conductors, splitting can be described by cubic Dresselhaus field away from zone center \cite{Dresselhaus}. The value of Dresselhaus coupling for GaAs lies between 9 to 28 eV \AA$^{3}$. Experimentally, it is difficult to determine these parameters however, theoretically accurate calculation of electronic band structure is required \cite{FabianMatos}. The incorporation of SO coupling improves the description of band structure and gives rise to many interesting phenomena in semi-conductors such as spin relaxation \cite{FabianDSarma}, spin orientation (optical) \cite{Meier}, spin Hall effect \cite{Sinova} and spin galvanic phenomena \cite{Ganichev} etc.\\
The electronic properties of these semi-conductors have been studied experimentally \cite{Fern,Adachi,Joyce,Garriga,Djurisic} and theoretically \cite{Remediakis,Lebegue,Koller,Tomic,Shimazaki,Kim,Wang,Hinuma} by several authors however, in all calculations within the formulism of density functional theory (DFT), band gap is underestimated. The strongly underestimated fundamental band gap in GaAs \cite{JP} yields spin-splitting parameter ($\Delta_{SO}$) 14 times larger in comparison to the value predicted by GW band structure \cite{Chantis}. To obtain a band gap that is consistent with experiment, it is necessary to acquire complete information about band structure including spin-orbit coupling \cite{Cardona,Winkler,Chantis}. Further, spin-orbit split-off parameter should be carefully calculated and this can be achieved by using phenomenological approach of \textit{\textbf{k $\cdot$ p}} method \cite{Kane,Sipahi}. In \textit{\textbf{k $\cdot$ p}} method, Hamiltonian is constructed by employing perturbation theory [28,29] %\cite{Enderlein,Willatzen}
and group theory analysis to reduce number of matrix elements which are replaced by effective parameters. The number of these parameters rely on the number of selected bands and on the symmetry of the described crystal. In ZB crystals, effective masses themselves allow the calculation of effective mass parameters \cite{Leite}. Theoretically, effective masses can be determined by fitting parabolic dispersion near $\Gamma$ point in the band structure \cite{Dugdale,Ramos}.\\
In this paper, we studied structural and electronic properties of ZB cubic phase GaAs in plane-wave density-functional theory (DFT) code JDFTx, by employing Becke 86 in 2D and PBE exchange correlation functionals from libXC. We included spin-orbit coupling by using full relativistic norm-conserving pseudopotentials and further applied \textbf{\textit{k $\cdot$ p}} method to fit conduction and valence bands to calculate the effective masses of non-degenerate and degenerate bands. Our results compare fairly well to existing findings and provide useful data for more empirical approaches.\\ The paper is organized as follows. Crystal structure of GaAs is shown in Section 2. Computational method is given in Section 3. Results are presented in Section 4 with discussion and comparison. The last section concludes our work.
\section{Crystal structure}
\begin{figure}[H]
\begin{center}
\includegraphics[width=0.6\linewidth]{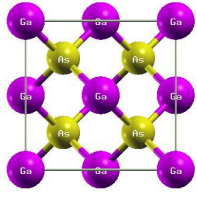}
\caption{Crystal structure of F-43M zinc-blende (ZB) 3D-cubic GaAs.}
\label{GaAs_structure}
\end{center}
\end{figure}

GaAs is a binary alloy of III-V semi-conductor group. The atoms arranged in zinc-blende (ZB) structure have tetrahedral coordination with two interpenetrating face-centered Bravais lattices, each with different atomic species. One specie is cation while other is anion. The particular order of cations and anions within unit cell determines spin orientation \cite{Cardona} caused by SO field. The crystal structure of F-43M ZB GaAs is shown in Fig. \ref{GaAs_structure} with Ga (cation) atoms positioned at origin and As (anion) atoms located at fractional coordinate (1/4,1/4,1/4) in a cubic lattice of length \textbf{a}.
\section{Computational method}
The exchange-correlation energy of electrons was represented by Becke 86 in 2D \cite{Becke86} and Perdew, Burke \& Ernzerhof \cite{PBE} from libXC \cite{libxc} in JDFTx \cite{JDFTx}. The valence electrons-ions interactions were represented by optimized norm-conserving Vanderbilt pseudopotentials (ONCVPSP) \cite{Hamann} in UPF format \cite{Pdojo}. We performed calculations with no-spin and relativistic spin-type to calculate structural and electronic properties of GaAs. The strain tensor was minimized to determine optimized lattice constant and after complete convergence study, energy cutoff of 50 Ha and k-points grid \cite{HJMonkhorst,JDPack} of 20 $\times$ 20 $\times$ 20 that was less than 0.0001 mHa converged, were selected. The optimized lattice constant was used to calculate band structure with and without SO coupling.
\section{Results and discussion}
\subsection{Structural properties}
The optimized lattice constant calculated by employing L-BFGS scheme \cite{LBFGS} was 5.711 \AA\hphantom{0}that is 1 \% off from experimentally determined value of 5.654 \AA\hphantom{0}which is quite reasonable when PBE exchange correlation functional is used. Although optimized lattice constant is 1\% off from experiment however, it is in satisfactory agreement with earlier reported findings. The comparison of our calculated value with earlier published theoretical and experimental values is given in Table ~\ref{parameter_compare}.
\begin{table}[H]
\centering
\caption{Comparison of our calculated lattice parameter with reported findings. The lattice constant is in units of \AA.}
\begin{tabular}{llll}
%\Hline % To generate a thicker line than \hline
\hline
Parameter & This work & Theoretical & Exp\\
%\Hline
\hline
a & 5.711 & 5.508$^{a}$, 5.651$^{b}$  & 5.654$^{e}$ \\
  &       & 5.726$^{c}$, 5.755$^{d}$  & 5.653$^{f}$ \\
%\Hline
\hline
\end{tabular}
\label{parameter_compare}
\center{$^{a}$ref.\cite{Agarwal}, $^{b}$ref.\cite{Min}, $^{c}$ref.\cite{Staroverov}, $^{d}$ref.\cite{Kalvoda}, $^{e}$ref.\cite{Vurgaftman}, $^{f}$ref.\cite{Fillipi}}
\end{table}
\subsection{Band structure}
\subsubsection{No spin-orbit}
\begin{figure}[htp]
\hspace*{\fill}%
\begin{minipage}[t]{0.5\textwidth}
\centering
\vspace{0pt}
\includegraphics[width=0.7\linewidth,angle=-90]{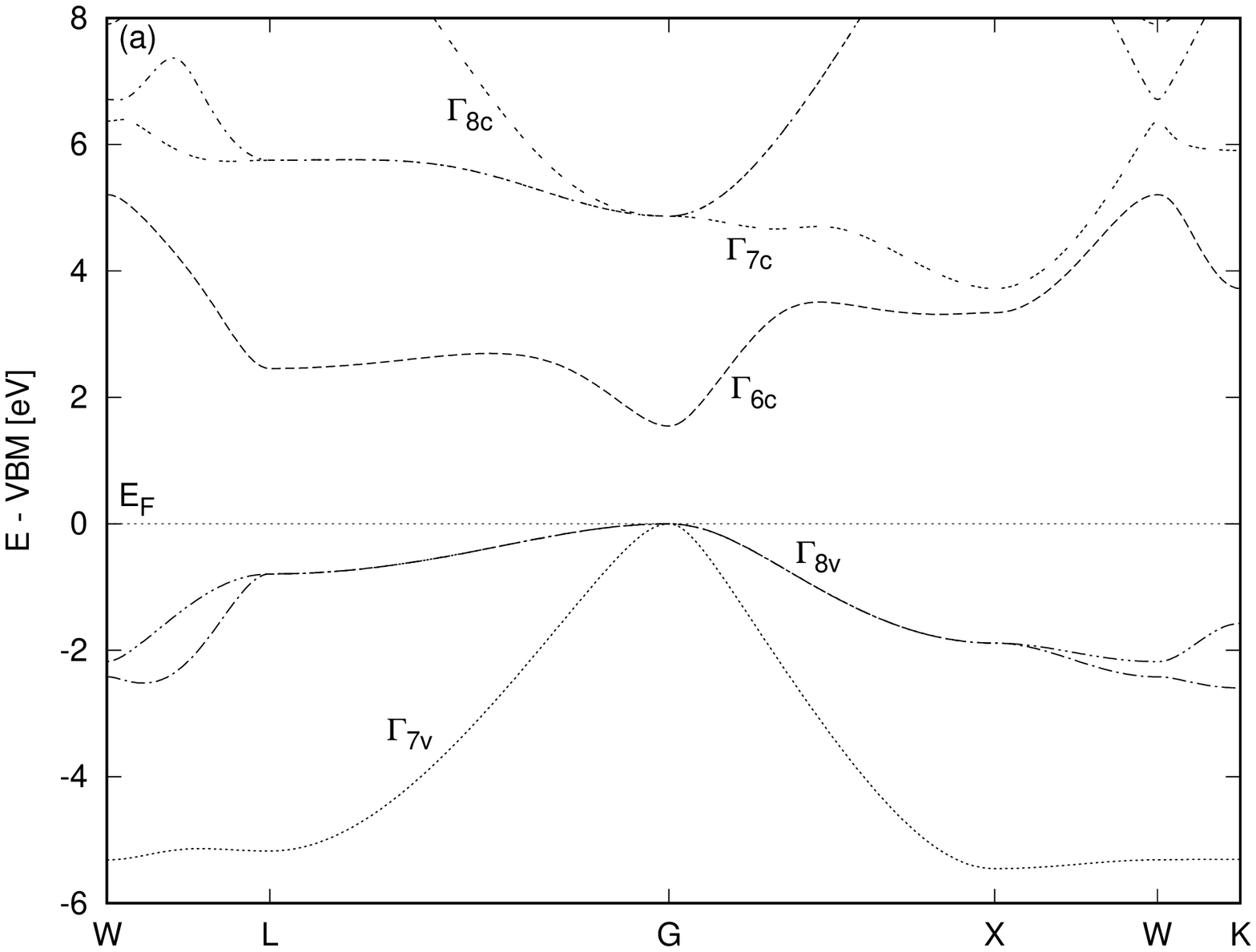}
\end{minipage}%
\hfill
\begin{minipage}[b]{0.5\textwidth}
\centering
\vspace{0pt}
\hspace{-1pt}
\includegraphics[width=0.7\linewidth,angle=-90]{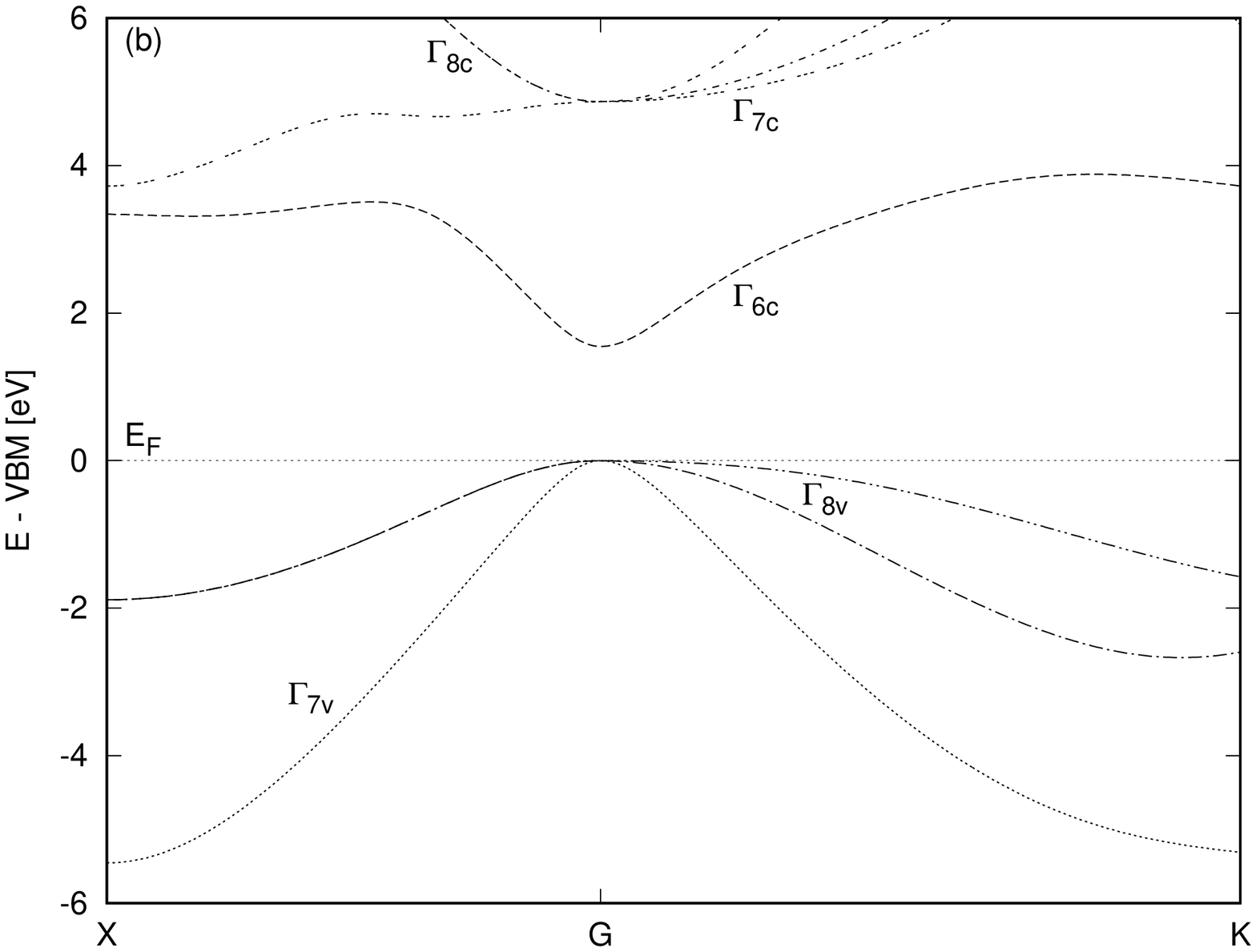}
\end{minipage}%
\hspace*{\fill}
\caption{Bands dispersion of GaAs without spin-orbit (SO) coupling. (a) along path W-L-G-X-W-K and (b) along path X-G-K. The Fermi level is shown as E$_{F}$.}
\label{bs_wo_so}
\end{figure}

Most of the density functional theory softwares underestimate lattice constant however, this was not the case with JDFTx therefore, we used it to calculate the band structure. The bands dispersion of GaAs was calculated without spin-orbit coupling along high symmetry path W-L-G-X-W-K and X-G-K using a primitive cell of two atoms. The symmetry point G represents zone center $\Gamma$. The direct band gap (E[$\Gamma_{6c}$]-E[$\Gamma_{8v}$]) between conduction band $\Gamma_{6c}$ and valence band $\Gamma_{8v}$ at zone center $\Gamma$ was 1.5480 eV. The gap of L-valley was 2.4569 eV and the gap of X-valley was 3.3398 eV. The gap (E[$\Gamma_{8c}$]-E[$\Gamma_{8v}$]) between conduction band $\Gamma_{8c}$ and valence band $\Gamma_{8v}$ was 4.8660 eV. The gap (E[$\Gamma_{6c}$]-E[$\Gamma_{6v}$]) between conduction band $\Gamma_{6c}$ and valence band $\Gamma_{6v}$ was 13.5645 eV. The gap (E[$\Gamma_{6v}$]-E[$\Gamma_{8v}$]) between valence band $\Gamma_{6v}$ and valence band $\Gamma_{8v}$ was -12.0165 eV. The bands dispersion without spin-orbit coupling along symmetry points W-L-G-X-W-K and X-G-K is shown in Fig. \ref{bs_wo_so} (a) and (b) respectively. The calculated band gap is in fair agreement with recently reported finding \cite{Ali}.
%\begin{figure}[!htb]
%\centering
%\begin{minipage}[t]{.5\textwidth}
%\centering
%\includegraphics[width=0.7\linewidth,angle=-90]{BS_WLGXWK.eps}
%%\caption{}
%%\label{fig:prob1_6_2}
%\end{minipage}%
%\begin{minipage}[b]{0.5\textwidth}
%\centering
%\includegraphics[width=0.7\linewidth,angle=-90]{BS_XGK.eps}
%%\caption{$dt =$}
%%\label{fig:prob1_6_1}
%\end{minipage}
%\caption{Bands dispersion of GaAs without spin-orbit (SO) coupling. (a) along path W-L-G-X-W-K and (b) along path X-G-K. The Fermi level is shown as E$_{F}$.}
%\label{bs_wo_so}
%\end{figure}
\subsubsection{Spin-orbit}
\begin{figure}[htp]
\hspace*{\fill}%
\begin{minipage}[t]{0.5\textwidth}
\centering
\vspace{0pt}
\includegraphics[width=0.75\linewidth,angle=-90]{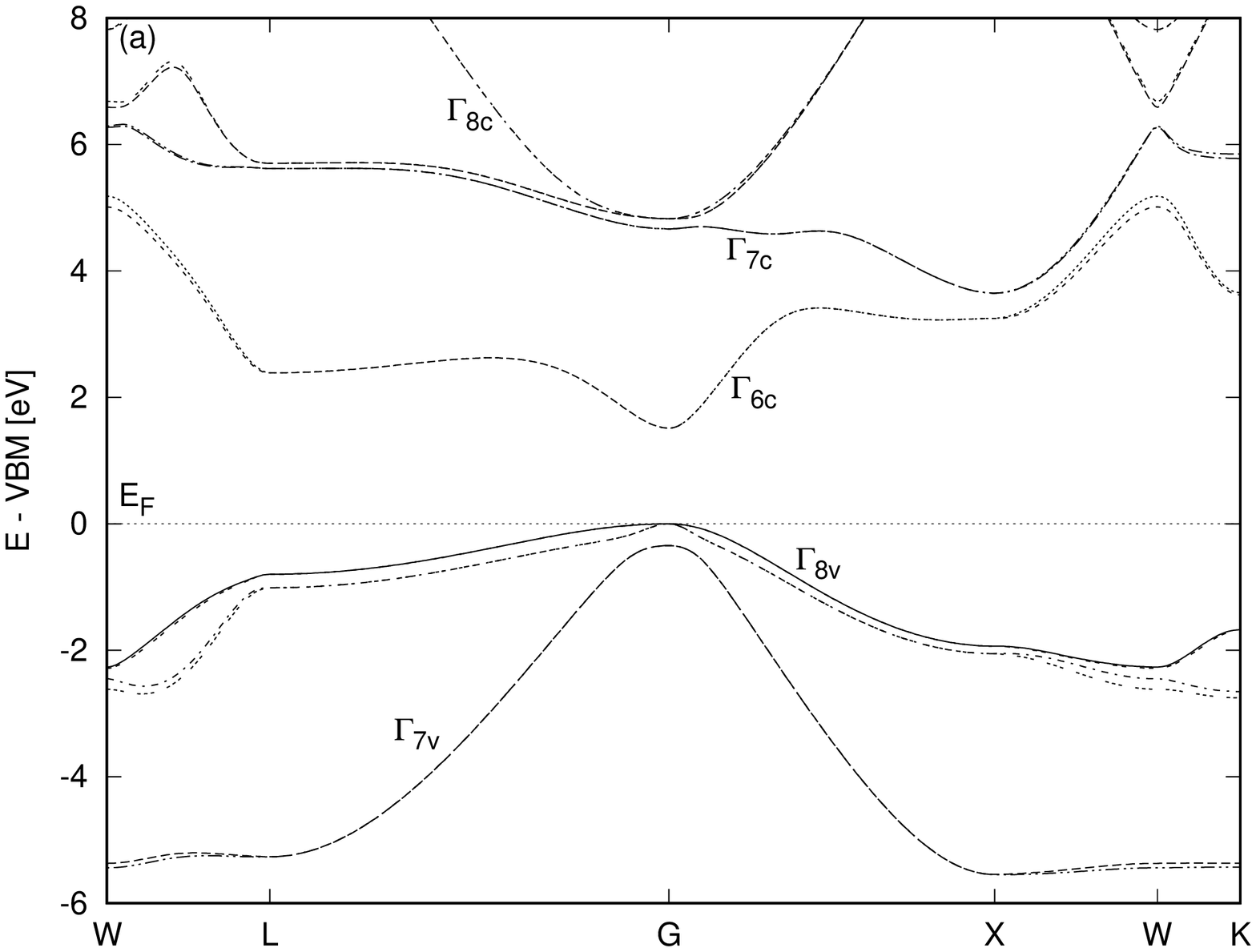}
\end{minipage}%
\hfill
\begin{minipage}[b]{0.5\textwidth}
\centering
\vspace{0pt}
\hspace{1.5pt}
\includegraphics[width=0.75\linewidth,angle=-90]{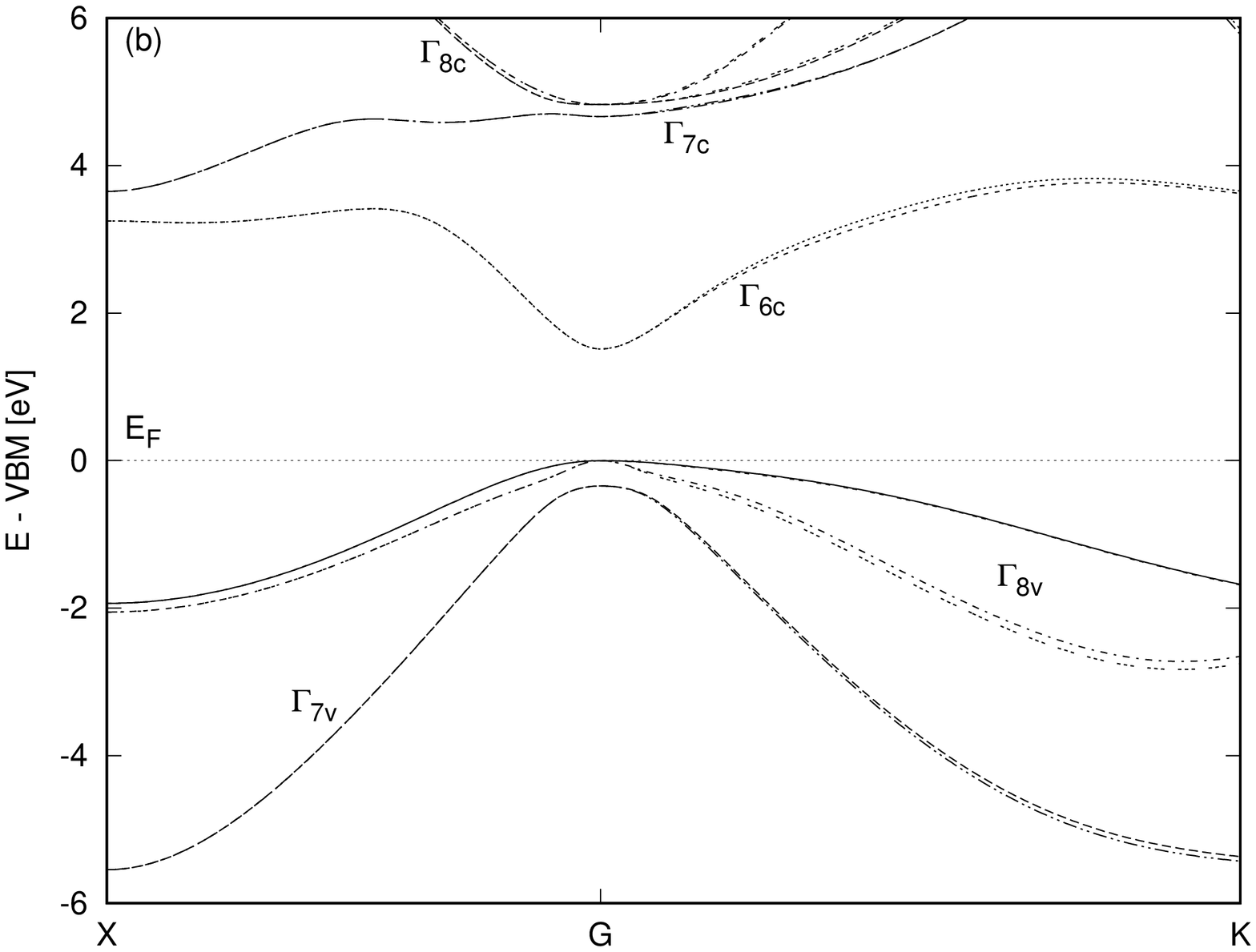}
\end{minipage}%
\hspace*{\fill}
\caption{Bands dispersion of GaAs with spin-orbit (SO) coupling. (a) along path W-L-G-X-W-K and (b) along path X-G-K. The Fermi level is shown as E$_{F}$.}
\label{bs_w_so}
\end{figure}

The SO coupling was included by incorporating full relativistic optimized norm-conserving Vanderbilt pseudopotentials. The band structure was calculated along high symmetry path W-L-G-X-W-K and X-G-K. The direct band gap (E[$\Gamma_{6c}$]-E[$\Gamma_{8v}$]) between conduction band $\Gamma_{6c}$ and valence band $\Gamma_{8v}$ at zone center $\Gamma$ was 1.5142  eV. The gap of L-valley was 2.3873 eV and the gap of X-valley was 3.2496 eV. The gap (E[$\Gamma_{8c}$]-E[$\Gamma_{8v}$]) between conduction band $\Gamma_{8c}$ and valence band $\Gamma_{8v}$ was 4.8272 eV. The gap (E[$\Gamma_{6c}$]-E[$\Gamma_{6v}$]) between conduction band $\Gamma_{6c}$ and the valence band $\Gamma_{6v}$ was 13.5529 eV. The gap (E[$\Gamma_{6v}$]-E[$\Gamma_{8v}$]) between valence band $\Gamma_{6v}$ and valence band $\Gamma_{8v}$ was -12.0387 eV. The spin-orbit split-off parameter was 0.3440 eV. The bands dispersion with spin-orbit coupling along high symmetry points W-L-G-X-W-K and X-G-K is shown in Fig. \ref{bs_w_so} (a) and (b) respectively. The direct manifestation of SO coupling in band structure is splitting of valence energy bands into light-hole (LH), heavy-hole (HH) and spin-orbit (SO) split-off hole bands. Further, the calculated values using JDFTx are not overestimated but in good agreement with the recently reported finding \cite{Ali}. 
%\begin{figure}[H]
%\centering
%\begin{minipage}{.5\textwidth}
%\centering
%\includegraphics[width=0.75\linewidth,angle=-90]{BS_SO_WLGXWK.eps}
%%\caption{}
%%\label{fig:prob1_6_2}
%\end{minipage}%
%\begin{minipage}{0.5\textwidth}
%\centering
%\includegraphics[width=0.75\linewidth,angle=-90]{BS_SO_XGK.eps}
%%\caption{$dt =$}
%%\label{fig:prob1_6_1}
%\end{minipage}
%\caption{Bands dispersion of GaAs with spin-orbit (SO) coupling. (a) along path W-L-G-X-W-K and (b) along path X-G-K. The Fermi level is shown as E$_{F}$.}
%\label{bs_w_so}
%\end{figure}
\subsection{Effective mass from \textit{\textbf{k}} $\cdot$ \textit{\textbf{p}} parameters}
The effective mass of the non-degenerate band can be calculated by using \textit{\textbf{k}} $\cdot$ \textit{\textbf{p}} method. The complete information of the method is given in ref. \cite{YuCordona}. By employing standard non-degenerate perturbation theory, eigenfunctions \textit{u$_{nk}$} and eigenvalues \textit{E$_{nk}$} at a neighbouring point \textit{\textbf{k}} are expanded up to second order in $k$ in terms of unperturbed wave functions \textit{u$_{n0}$} and energies \textit{E$_{n0}$}. Let's consider the lowest conduction band (CB) at the zone center that has symmetry $\Gamma_{1}$. According to \textit{\textbf{k}} $\cdot$ \textit{\textbf{p}} method, the effective mass can be determined mainly through its coupling to the nearest bands with $\Gamma_{4}$ symmetry, via the \textit{\textbf{k}} $\cdot$ \textit{\textbf{p}} term. These bands include both conduction and valence bands. The momentum matrix element between $\Gamma_{1}$ conduction band and $\Gamma_{4}$ conduction band is smaller in comparison to the momentum matrix element between $\Gamma_{1}$ conduction band and $\Gamma_{4}$ valence band. Therefore, for III-V group semiconductors, under the condition of parabolicity at the extreme band edge of the conduction band, we have
\begin{equation}\label{effmass}
\frac{m}{m_c^*} \approx \frac{2 P^2}{m E_0} \ ,
\end{equation}
where \textit{$E_0$} is the direct separation between conduction band (CB) and valence band (VB) i.e. the band gap of the material and the matrix element $P^2$ is approximately same for III-V semiconductors with $2P^2/m \approx$ 20eV \cite{YuCordona}. The CB band minimum at zone center satisfies the condition of parabolicity and its minimum is the minimum of parabola fit within 5 \% of FBZ. Therefore, the conduction band effective mass (m$_{c}^*$) was calculated by using Eq. (\ref{effmass}). The calculated effective mass of conduction band (m$_{c}^*$) was $0.0525 m$, that is in fair agreement with the reported value in ref. \cite{YuCordona}.\\
To calculate the effective mass of the degenerate band at the valence band extremum at the zone centre, we used 6 $\times$ 6 zinc-blende (ZB) effective \textit{\textbf{k}} $\cdot$ \textit{\textbf{p}} Hamiltonians proposed by Luttinger and Kohn (LK6) \cite{LuttingerKohn} and extended by Kane \cite{Kane} to a 8 $\times$ 8 model. According to LK6 model, class A is composed of three top most valence bands such as heavy-hole (HH), light-hole (LH) and spin-orbit split-off (SO) hole near $\Gamma$ point. Kane included first conduction band (CB) in class A by using same model and perturbative order. The Kane Hamiltonian depends on five different effective mass parameters such as $\tilde{\gamma_{1}}$, $\tilde{\gamma_{2}}$, $\tilde{\gamma_{3}}$, $\tilde e$ and $P$ along with band gap and $\Delta_{SO}$. However, in LK6 model, three parameters such as $\gamma_{1}$, $\gamma_{2}$ and $\gamma_{3}$ along with $\Delta_{SO}$ are important \cite{Enderlein,Willatzen}. The authors in refs. \cite{PRB,control} also used k $\cdot$ p method and more recently Bastos et al. \cite{Bastos} used electronic g-factors that are directly linked to the spin splitting of the carrier bands, to calculate the effective mass by using the Roth's formula \cite{Roth} given by
\begin{equation}
g_{c}^* = 2 - \frac{2 E_{P} \Delta_{SO}}{3 E_{gap}(E_{gap} + \Delta_{SO})} \ ,
\end{equation}
and the values of $E_P$, E$_{gap}$ and $\Delta_{SO}$. The equation given above only considers interaction between valence band (VB) and conduction band. The interactions between other bands are neglected. The relations between both model parameters are given by
\begin{equation}
\begin{split}
&\gamma_{1} = \tilde{\gamma_{1}} + \frac{E_p}{3 E_{gap} } \ , \ \ \ \gamma_{2} = \tilde{\gamma_{2}} + \frac{E_p}{6 E_{gap}} \ , \ \ \ \gamma_{3} = \tilde{\gamma_{3}} + \frac{E_p}{6 E_{gap}} \ , \\& e = \tilde{e} + \frac{\left(E_{gap} + \frac{2}{3} \Delta_{SO}\right)E_p}{E_{gap}\left(E_{gap} + \Delta_{SO}\right)} \ , \ \ \ E_p = \frac{2m_0}{\hbar^2} P^2 \ .
\end{split}
\end{equation}
The effective masses as determined from these parameters are given by the relations
\begin{equation}
\begin{split}
&m_{lh}[100]  = (\gamma_{1} + 2 \gamma_{2} )^{-1} \ , \ \ \ m_{lh}[110]  = (\gamma_{1} + 2 \gamma_{3} )^{-1} \ , \\& m_{hh}[100]  = (\gamma_{1} - 2 \gamma_{2} )^{-1} \ , \ \ \ m_{hh}[110]  = (\gamma_{1} - 2 \gamma_{3})^{-1} \ , \\& m_{lh}[111]  = (\gamma_{1} + \sqrt(\gamma_{2}^2 + 3 \gamma_{3}^2)^{-1} \ , \ \ \ m_{e} = e^{-1} \ , \\& m_{hh}[111]  = (\gamma_{1} - \sqrt(\gamma_{2}^2 + 3 \gamma_{3}^2)^{-1} \ , \\& m_{SO} = \left(\gamma_{1} - \frac{1}{3} \frac{\Delta_{SO} E_{p}}{E_{gap} (E_{gap} + \Delta_{SO})} \right)^{-1} \ .
\end{split}
\end{equation}
The effective masses of degenerate valence bands along [100], [110] and [111] direction were calculated by using the aforementioned relationships and the results are tabulated in Table \ref{effmasstable}. For the comparison, the results of Bastos et al. \cite{Bastos} are given. Our calculated value of $\Delta_{SO}$ was 0.34 eV that is in satisfactory agreement with the value reported in ref. \cite{Bastos,Madelung}.
\begin{table}[H]
\begin{center}
\caption{Comparison of our calculated values with Bastos et al.'s results.}
\label{t1}
\begin{tabular}{lll}
%\Hline % To generate a thicker line than \hline
\hline
Parameter & This work & Ref.\cite{Bastos} \\
%\Hline
\hline
E$_{p}$ & 29.1250 eV & 25.9-27.6 \\
$\gamma_{1}$ & 6.850 & 6.8-7.8\\
$\gamma_{2}$ & 2.060 & 2.02-2.5\\
$\gamma_{3}$ & 2.420 & 1.0-2.43\\
m$_{lh[100]}$ & 0.0912 & 0.088 \\
m$_{lh[110]}$ & 0.0855 & 0.079 \\
m$_{hh[100]}$ & 0.3663 & 0.357 \\
m$_{hh[110]}$ & 0.4975 & 0.672 \\
m$_{lh[111]}$ & 0.0868 & 0.076 \\
m$_{hh[111]}$ & 0.4588 & 0.898 \\
m$_{SO}$ & 0.1766 & 0.169 \\
$g_{c}^*$ & -0.37  & -0.34 \\
$\Delta_{SO}$ & 0.3440 & 0.325-0.365 \\
%\Hline
\hline
\label{effmasstable}
\end{tabular}
\end{center}
\end{table}
\section{Conclusion}
We studied structural and electronic properties of zinc blende (ZB) cubic phase GaAs by employing Becke 86 in 2D and PBE exchange correlation functionals. Several authors have reported theoretical results on GaAs by using GGA and ultrasoft pseudopotentials with underestimated band gap and overestimated spin-splitting parameter however, we used standard ONCVPSP in plane-wave joint density-functional theory code. The optimized lattice constant of 5.711 \AA\hphantom{0}obtained using L-BFGS algorithm is 1\% off from experimental value of 5.654 \AA\cite{Vurgaftman} and it is in fair agreement with theoretical value of 5.742 \AA\cite{Bastos}. The band gap of 1.5142 eV with SO coupling and spin-orbit split-off parameter are in excellent agreement with experiment \cite{Bastos,Madelung}. The effective mass parameters calculated using \textit{\textbf{k $\cdot$ p}} method are in fair agreement with most recent findings of Bastos et al. \cite{Bastos}. Our findings provide useful data for further empirical studies related to semiconductor spintronic devices. Lastly, we have used open source code JDFTx in our calcualtions that is recently introducted and we predict that its accuracy is quite reasonable.
\section*{Acknowledgement}
This work was supported by the National Science Foundation of China under Grant No.11674223.
%%References section

\end{document}